\documentclass[twocolumn,aps]{revtex4}
\usepackage[dvips]{graphicx}
\usepackage{mathrsfs}
\usepackage{amsmath}
\usepackage{dcolumn}

\begin{document}
\title{Ripple Distribution for Nonlinear Fibre-Optic Channels}
\author{Mariia Sorokina$^{*}$, Stylianos Sygletos,  and Sergei Turitsyn }

\address{Aston Institute of Photonic Technologies, Aston University, B4 7ET Birmingham UK}

\email{m.sorokina@aston.ac.uk } 

\begin{abstract}
Since Shannon proved that Gaussian distribution is the optimum for a linear channel with additive white Gaussian noise and he calculated the corresponding channel capacity, it remains the most applied distribution in optical communications while the capacity result is celebrated as the seminal “linear Shannon limit”. Yet, when it is applied in nonlinear channels (e.g. fiber-optics) it has been shown to be non-optimum, yielding the same result as for uncoded transmission in the high nonlinear regime.  This has led to the notion of "nonlinear Shannon limit", which predicts vanishing capacity at high nonlinearity. However, recent findings indicate that non-Gaussian distribution may lead to improved capacity estimations, urging for an exciting search for novel methods in nonlinear optical communications.  Here for the first time, we show that it is possible to transmit information above the existing limits by using a novel probabilistic shaping of the input signal, which we call ripple distribution
\end{abstract}
\maketitle
Fiber-optic transmission is the back-bone of communication systems, yet its capacity -- maximum error-free transmission rate -- remains unknown due to intricate non-linear memory effects.
As Shannon found optimum distribution for a linear additive white Gaussian noise channel to be Gaussian distribution\cite{Shannon}, it is widely applied in nonlinear fiber-optic systems\cite{Sp,Essiambre,GN}. Since then, a number of coding schemes and modulation formats \cite{IPM} have been proposed, creating a family of probabilistic shaping based on Gaussian distribution with the maximum increase of the order of $0.2$ bits compared to uncoded transmission \cite{GS1,GS3,GS4}.

However, the Gaussian shaping assumption provides data rate estimates below the nonlinear threshold (also referred as "nonlinear Shannon limit")\cite{DJR,Radic1,PW,chapter,Agrell2015}, which are also overly pessimistic, as it as been recently pointed out in \cite{Agrell2014}. This is because the derived models make averaging of the signal dynamics and lose information about inter-symbol interference effects. Other widely used practical approaches, the so-called "perturbative models with deterministic nonlinearity"\cite{D1,D3,Dar,Tao}, achieve a first-order perturbative solution of the nonlinear Schr\"{o}dinger equation by taking into account only the signal-signal interactions in fiber transmission. Such type of nonlinear signal distortion is deterministic and it can be compensated, in principle, with some elaborate technical efforts\cite{Radic1}. Thus, the principal challenge is to provide an accurate analysis of the signal-noise interactions with signal-dependent statistics.

Here, for the first time we achieve data rates above the conventional limits  by introducing a novel type of input signal distributions  -- ripple distributions. Moreover, the results demonstrate that monotonically increasing transmission rates can be achieved even in the high nonlinear regime. These findings are in sharp contrast to previous estimates based on the Gaussian distribution of input signal and break the notion of "capacity vanishing to zero at high signal power" thus establishing a new direction for coding and signal channel distortion compensation algorithms.
\\ \\
\textbf{Finite Memory Discrete-Time Channel Model.} A typical communication system is presented in Fig. 1a). At the transmitter, the message (uncoded bits) is modulated to a discrete time set of constellation symbols (here, 64-QAM plotted in Fig. 1b) and after pulse shaping it is mapped to a continuous time signal (Fig. 1c), which is subsequently launched to a multi-span optical fiber link.
The propagation of the continuous-time signal $E(t,z)$ in the optical fiber (Fig. 1d) is governed by the well known nonlinear Schr\"{o}dinger equation (NLSE):
\begin{equation}\label{NLSE}
\frac{\partial E}{\partial z}=-\frac{\alpha}{2}E-i \frac{\beta_2}{2}
\frac{\partial^{2} E}{\partial^{2} t}+ i\gamma|E|^2 E +
\eta(t,z),
\end{equation}
where the deterministic distortions (Fig. 1e) are introduced by fiber loss $\alpha$, second-order dispersion parameter $\beta_2$, and by Kerr nonlinearity characterized by the coefficient $\gamma$. The stochastic
distortions are described by the zero-mean additive white Gaussian noise (AWGN)  $ \eta(t,z)$ of variance $\langle \eta(z,t),\eta^*(z',t') \rangle=\frac{D}{L}\delta(z-z')\delta(t-t')$, with $D$ and $L$ being the
noise spectral density and the transmission length, respectively.

A discrete-time channel model is crucial for information-theory based analysis as it enables optimization of the mutual information functional $I(\mathbf{X},\mathbf{Y})$ for deriving the optimum signal distribution  $P(\mathbf{X})$ (Fig. 1f) and calculating the maximum reliable transmission rate, which is the channel capacity $C=\max I(\mathbf{X},\mathbf{Y})$. The transition between continuous-time modeling, given by Eq.\ref{NLSE}, to discrete-time modeling is not straightforward, since it requires expansion over a complete orthogonal set of basis functions $\{f_k(t)\}$. This is equivalent to matched filter demodulation at the receiver for generating observable discrete-time variables $\{Y_k\}$ (Fig. 1g). At the transmitter, signal expansion over the carrier pulses is considered, that is $E(t,0)=\sum_{k=-\infty}^{\infty}X_kf(t-kT)$, where $X_k$ are the complex modulated symbols, $f(t)$ is the time-varying pulse waveform and $T$ is the symbol period. At the receiver, the signal undergoes matched filtering, dispersion compensation and sampling at $t=kT$: $Y_k(\xi)=P^{-1/2}\int dt \mathrm{D}[E(t,\xi)]f(t-kT)$ (with dimensionless coordinate $\xi=z/L_d$ and $L_d$ denotes the dispersion length), which allows the following discrete-time representation of NLSE:
\begin{equation}\label{DT-NLSE}
Y_{k}'=\frac{L_d\rho}{L}\eta_k+\varepsilon V[Y]_k, \end{equation}
\[  V[Y]_k= \Psi_s(\xi)\sum_{m,n=-M}^{M}
Y_{k+m}(\xi)Y_{k+n}(\xi)Y^*_{k+m+n}(\xi)\tilde{C}_{mn}(\xi)\]
where the AWGN noise term $ \eta_k$ is characterized by the correlation $\langle \eta_k(\xi),\eta_{k'}^*(\xi') \rangle=\Psi_n(\xi)\delta(\xi-\xi')\delta_{kk'}$,
where $\Psi_n(\xi)=\lfloor \xi/Ls\rfloor^2 e^{-\alpha mod(\xi,Ls/Ld)}$ is the noise power profile and $\lfloor x\rfloor$ denotes floor function over variable $x$. Also $\Psi_s(\xi)=e^{-\alpha mod(\xi,L_s/L_d)}$
represents the signal power profile and $L_s$ is the span length. The coupling coefficients $\tilde{C}_{mn}$ define the memory behavior (within a memory window $M$) of the transmission channel and depend on its
physical properties and the signal pulseshape:

\begin{figure*}
\begin{center}
\includegraphics[width=17cm]{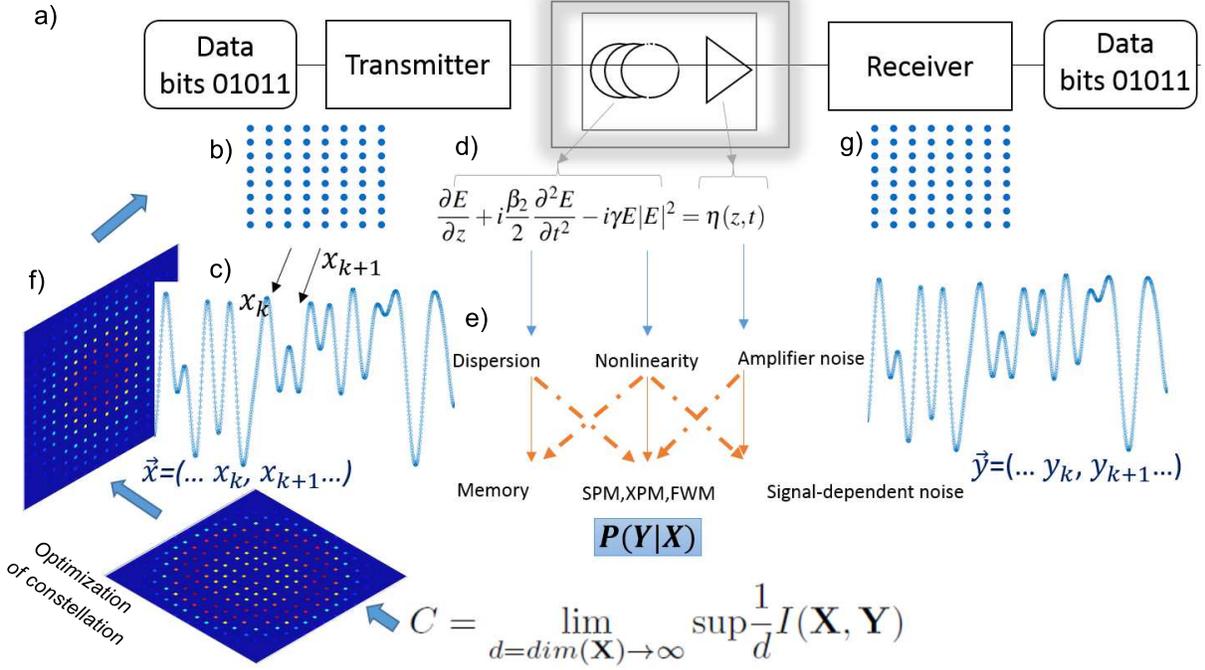}
\end{center}
\caption{  \label{F_C}  \textbf{Fiber-optic communication system.} a) The fundamental building blocks of a communication system where data is  coded to a discrete set of symbols $X_k$ (panel b) and transformed into continuous time form $E(t,0)$ (panel c) to be transmitted via the fiber channel. During transmission the signal is governed by the NLSE (panel d), which results in distortions: dispersion, nonlinearity, noise (panel e), which are reflected in the conditional pdf $P(\mathbf{Y}|\mathbf{X})$, which we can use to optimize constellation and coding  (panel f), to receive the maximum achievable transmission rate -- channel capacity $C$.  The received signal $E(t,L)$ is processed and sampled $Y_k$ and, finally, decoded to receive the data (panel g).
  }
\end{figure*}

\begin{equation}\label{Cmn_z}
\tilde{C}_{mn}=i  \int \int \int  d\omega  d\omega_1 d\omega_2 e^{-i\omega_1\omega_2\beta_2L_d \xi-i\omega_1mT-i\omega_2nT} \times \end{equation}
\[f^*(\omega)f(\omega_1+\omega)f(\omega_2+\omega)f^*(\omega_1+\omega_2+\omega)\]

\begin{figure*}
\begin{center}
\includegraphics[width=17cm]{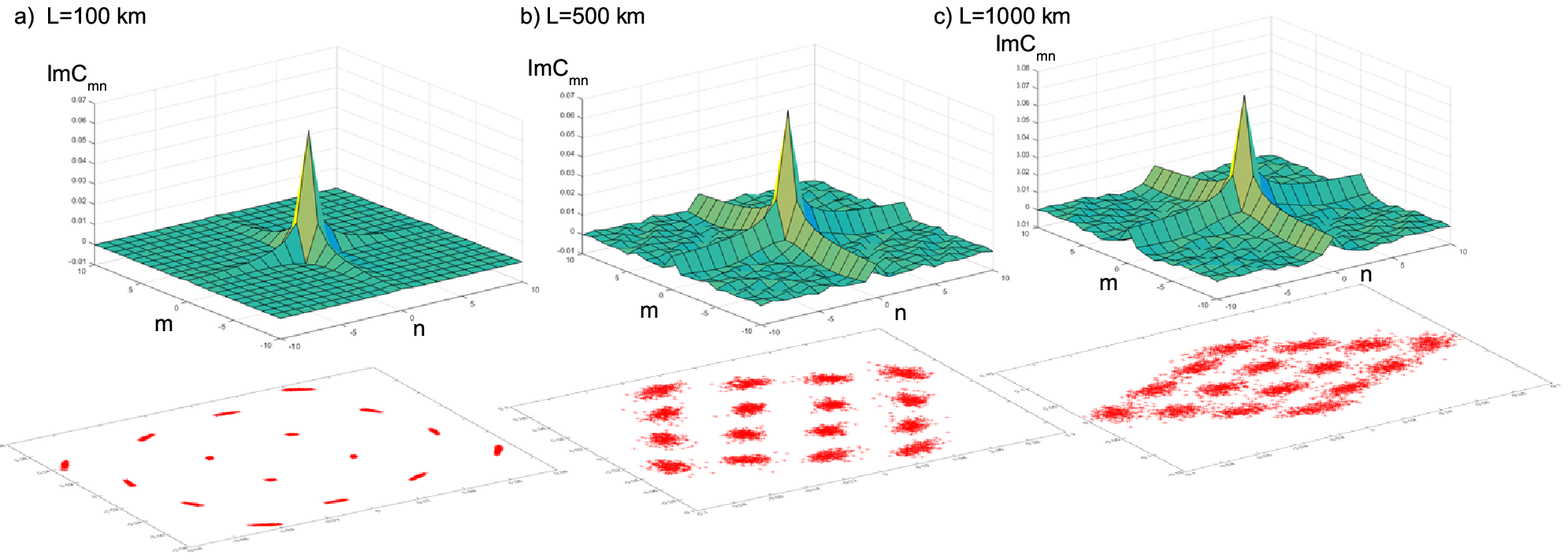}
\end{center}
\caption{  \label{Cmn} \textbf{Coupling matrix}. For various transmission distances a) $L=100 \ km$, b) $L=500 \ km$, c) $L=1000 \ km$ and the corresponding received constellation diagrams below (for input power
6dBm and span length 100km). We see that the coupling matrix reflects the strength of the inter symbol interaction and its effect on the signal distortion.}
\end{figure*}

The proposed channel model generalizes a number of previous results. Namely, the class of  infinite-memory Gaussian noise models, reported in \cite{Sp,Essiambre,GN,Karlsson}, can be received from Eq.
\ref{MCM1} after averaging the nonlinear interference term $\langle|C_{mn}|^2\rangle S^3$, whereas the finite memory model of \cite{Agrell2014} can be received by averaging of the coupling matrix
$\langle|C_{mn}|^2\rangle$ while keeping the information about interfering symbols. Thus, this is the first finite-memory discrete-time model that:  a)	captures pulse-shape and format dependence; b) includes signal-signal and signal-noise interactions; and c) enables incorporation of higher-order terms and derivation of conditional pdfs, which are essential for accurate capacity estimations.

 We  also employ multiple-scale analysis over the two small parameters that characterize the main signal degradation effects, that is : a) nonlinearity $ \varepsilon= L_d/L_{NL} $ (with dispersion length $L_d=T_0^2/\beta_2$  and  nonlinearity length $L_{NL}=1/(\gamma P)$) and b) noise $\rho=\sqrt{N/P}=1/\sqrt{SNR}$ (which is the reversed square root of the signal-to-noise-ratio (SNR) in the corresponding linear system) with noise power $N=DB$ and $B$ denoting the signal bandwidth. In the main order we have a linear channel with the AWGN noise term $\zeta_k$ characterized by the correlation $\langle \zeta_k,\zeta^*_m\rangle=\delta_{km}$.

The discrete-time perturbative multivariate channel model has a tensor form:
\begin{equation}\label{MT}
\mathbf{Y}=\mathbf{\widetilde{X}}+\mathbf{M}\mathbf{\zeta}+\mathbf{L}\mathbf{\zeta}^*,  \ \ \ \zeta=\int d\xi \eta(\xi)
\end{equation}
which  can be rewritten as follows
\begin{equation}\label{MCM1}
Y_k=\widetilde{X}_k+\sum_{m} M_{k,m}\zeta_m+L_{k,m}\zeta^*_m
\end{equation}
here, the first term describes the deterministic Kerr-effect induced inter-symbol interference on the transmitted symbol $X_k$ in the $k$-th time slot. It can be calculated by solving the deterministic part of Eq.
\ref{DT-NLSE} and it can be compensated at the transmitter or the receiver\cite{Radic1}. By expanding over small parameter $\varepsilon$ we can receive
\begin{equation}\label{Xtilde}
\widetilde{X}_k=Y_k^{(0)}+\sum_{N_o=1}^{\infty}\varepsilon^{N_o} Y_k^{(N_o)}, \ \ \ Y_k^0=X_k
\end{equation}
\\
\\
\textbf{Coupling Matrix.} The coupling matrix $C_{mn}=\int dz \Psi_s(z) \tilde{C}_{mn}(z)$ governs signal-signal interactions, particularly those which are responsible for the non-circular distribution of the distortion. This is achieved by taking also into account the pulse-shape impact. Its elements represent weights of interference between symbols in different time slots. To demonstrate the effect we considered the transmission of a single-channel in a dispersion unmanaged fiber  link. For simplicity, we used Gaussian pulses of 10 ps full width at half maximum duration and  baudrate of 28 GBaud. The link parameters were $\alpha=0.2 dB/km, \ \beta_2=-20 ps^2/km, \gamma=1.3 \ 1/W/km, \ \  L_s=100km $. The multi-span modeling of $1^{st}$ order approximation was verified numerically and experimentally\cite{ECOC_Essiambre,quant,16QAM}.  Fig. \ref{Cmn} shows that the strength of interference between neighbouring symbols decays exponentially with their distance (denoted by $m,n$), whereas the slope is defined by the parameters of the transmission system. At small transmission distances Fig. \ref{Cmn}a) we  observe interference between the closest neighbours, which causes dominance of phase distortion (see constellation diagram below). As the distance increases, more symbols interact and we observe more circular clouds (Fig. \ref{Cmn}b,c). The coupling matrix accurately identifies the interacting symbols and the non-uniform strength of their interference. This is an extremely important characteristic, since by preserving all relevant information about the signal interaction, it allows complete compensation at the transceiver. The latter cannot be achieved with a conventional approach that is based on averaging the signal statistics \cite{Sp,Essiambre,GN,Karlsson}.

\textbf{Signal-Signal Interactions. } Firstly, in Fig. \ref{F_C}a), we calculate the achievable data rates on a nonlinear transmission system of 1000 km length (without any type of nonlinear compensation), using the existing GN-model \cite{GN} and compare them with our approach based on calculating the variance of nonlinear distortions in  Eq. \ref{MCM1} $N_{S-S}=\langle|X-\tilde{X}e^{-i\phi_{nl}}|^2\rangle=2S^3 \sum_{m,n\neq 0}|C_{mn}|^2$ (the limit on summation reflects compensation of the stationary phase shift $\phi_{nl}=2\langle|X|^2\rangle\sum_{m}|C_{m0}|^2$). We can see that our developed model converges to Gaussian noise model under infinite-memory approximation.

The deterministic distortions can be compensated with the traditional digital back propagation or pre-distortion methods, or alternatively, with the perturbation approach of our developed channel model.
The perturbation over small parameter $\varepsilon$ defines deterministic signal distortion $Y^{(N_o)}$ of order $N_o$  by the recurrence relation:
  \begin{equation}\label{Yk}
  Y_k^{(N_o)}=\sum_{\substack{i,j,k=0 \\ i+j+l=N_o-1}}^{N_o} \sum_{m,n} C_{mn} Y^{(i)}_{k+n}Y^{(j)}_{k+m}(Y^{(l)}_{k+n+m})^*
  \end{equation}
The discrete-time characteristic of the approach makes possible the compensation of the nonlinear interference at the receiver without increasing the signal bandwidth. This is in contrast to the traditional pre-distortion techniques that are based on continuous-time waveform processing.  After removing the deterministic nonlinear distortions, signal-noise interference becomes the main limitation for increasing the transmitted information rates. Our proposed approach represents the first accurate discrete-time channel model with memory, which can uniquely capture such signal-noise beating effects.
\\
\\
\textbf{Signal-Noise Interactions. }
  The next two terms in Eq. \ref{MCM1} determine stochastic effects. If $M_{k,m}=\delta_{0,m}$ and $L_{k,m}=0$ we receive a linear AWGN channel; in the next order over parameter $\varepsilon$ signal-noise mixing effects are taken into account
\begin{subequations}
\begin{align}
M_{k,m}=\rho \delta_{k,m} + \rho \varepsilon\sum_n K_{n,m-k}(\widetilde{X}_{k+n}\widetilde{X}_{m+n}^*+\widetilde{X}_{m+n}\widetilde{X}_{k+n}^*), \\
L_{k,m}=\rho \varepsilon\sum_n K_{n,m-k-n}\widetilde{X}_{k+n}\widetilde{X}_{m-n}
\end{align}
\end{subequations}
with the matrix $K_{mn}=\int dz \sqrt{\Psi_n(z) \Psi_s(z)} \tilde{C}_{mn}(z)$.

Moreover, the matrix model of Eq. \ref{MT} represents a general form that can be easily expanded to cover multi-wavelength operation.  In that case, Eq. \ref{MCM1} can be rewritten as follows:
\begin{equation}\label{1}
Y_{k\theta}=\widetilde{X}_{k\theta}+\sum_{m,\alpha} M_{k\theta m\alpha}\zeta_{m\alpha}+L_{k\theta m\alpha}\zeta^*_{m\alpha}
\end{equation}
here Greek and Latin letters denote frequency and time indexes respectively.
\\
\\
\textbf{Capacity Lower Bounds and Conditional pdf. }
To derive the capacity  it is necessary to optimize the mutual information functional:
 \begin{equation}\label{Gamma}
 C=\lim_{d=dim(\mathbf{X})\rightarrow\infty} \mathrm{sup}\frac{1}{d} I(\mathbf{X},\mathbf{Y})
  \end{equation}
  with power constraint: $\int d\mathbf{x} ||\mathbf{x}||^2 P_{\mathbf{x}}=dS$.
To optimize the mutual information over input pdf one must calculate  the multivariate conditional pdf that takes into account the memory effects defined by the channel model of Eq. \ref{MT}.
Next, we derive the \emph{conditional pdf} for the transmitted symbols.
The channel model of Eq. \ref{MT} represents a mixing of signal and noise components as a result of the intersymbol interference.
A linear combination of univariate independent and identically distributed normal vectors can be represented as a complex normal distribution\cite{complexnormal}
\begin{equation}\label{cpdf_all}
P(\mathbf{Y}|\mathbf{X})= (\pi)^{-d}(|\mathbf{\Gamma}||\mathbf{P}|)^{-1/2}
 \end{equation}
 \[
\exp\Big[-\frac{1}{2}[(\mathbf{y}-\mathbf{\widetilde{x}})^H, (\mathbf{y}-\mathbf{\widetilde{x}})^T]
 \begin{pmatrix}
  \mathbf{\Gamma} & \mathbf{\Upsilon} \\
  \mathbf{\Upsilon}^H & \mathbf{\Gamma}^*
 \end{pmatrix}^{-1}
 \begin{pmatrix}
  \mathbf{y} - \mathbf{\widetilde{x}} \\
  \mathbf{y}^* -\mathbf{\widetilde{x}}^*
 \end{pmatrix}
 \Big]
\]
\textit{Notations}: $^*$ means the complex conjugate, $^T$ means transposition, and $^H$ means  transposition and complex conjugate.
\\
The covariance matrix $\mathbf{\Gamma}$ (real, symmetric, non-negative definite and Hermitian) and relation matrix $\mathbf{\Upsilon}$ (real and symmetric) are given as
 \begin{equation}\label{Gamma_def}
 \mathbf{\Gamma}=\mathrm{E}[(\mathbf{Y}-\mathbf{\widetilde{X}})(\mathbf{Y}-\mathbf{\widetilde{X}})^H], \ \ \ \mathbf{\Upsilon} =\mathrm{E}[(\mathbf{Y}-\mathbf{\widetilde{X}})(\mathbf{Y}-\mathbf{\widetilde{X}})^T] \end{equation}
\[\mathbf{P} =\mathbf{\Gamma}^*-\mathbf{\Upsilon}^H\mathbf{\Gamma}^{-1}\mathbf{\Upsilon}\]

\begin{figure*}
\begin{center}
\includegraphics[width=17cm]{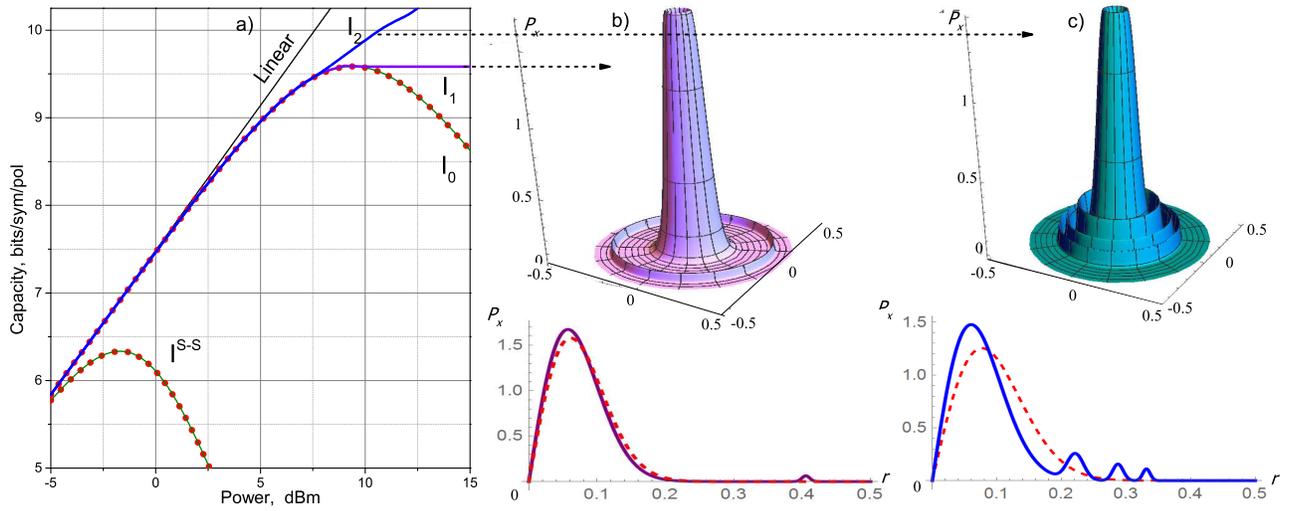}
\end{center}
\caption{  \label{F_C}  \textbf{Capacity and ripple distribution}. a) Capacity lower bounds for uncompensated signal-signal (S-S) distortions  $I^{S-S}$ (GN model (red, dotted) and the proposed model (green))  and
compensated deterministic distortions with account of signal-noise (S-N) interference. Previous lower bound (denoted by $I_0$) decreases to zero, whereas ripple distribution as input pdf (Eq. \ref{Px}) allows to
achieve higher monotonically increasing bounds, denoted by $I_1$ and $I_2$ for the increased number of ripples in an input pdf as shown in panels b) and c) correspondingly (Gaussian pdf is shown in red dashed lines
for comparison).}
\end{figure*}

Recalling Eq. \ref{MT}
 \begin{equation}\label{Gamma}
 \mathbf{\Gamma}= (\mathbf{M}\mathbf{M}^H+\mathbf{L}\mathbf{L}^H), \ \ \ \mathbf{\Upsilon} =(\mathbf{M}\mathbf{L}^T+\mathbf{L}\mathbf{M}^T)
  \end{equation}
This is the first result that presents the conditional pdf derived from a discrete-time model with memory that has been accurately defined by nonlinear properties of the fiber optic channel. This approach allows to capture any non-circular behaviour of signal distortions and it contains precise information about intersymbol and signal-noise interfering effect. In this particular case of $\mathbf{\Upsilon}=0$ and $E(\mathbf{Y})=0$ we have circularly symmetric complex normal distribution $\mathbf{Y}\sim \mathcal{CN}(0,\mathbf{\Gamma}$), i.e. $P(\mathbf{y})=(\pi)^{-d}|\mathbf{\Gamma}|^{-1}e^{-\mathbf{y}^{H}\mathbf{\Gamma}^{-1}\mathbf{y}}$.
\\
\\
\textbf{Ripple distribution. }
Now let us  consider a set of $\alpha=1..q$ Gaussian distributions with uniform phase, each of which is localized around a different power level $\rho^2_\alpha$ and it has different variance $S_\alpha$ and weight
$p_{\alpha}$, so that in polar coordinates  it is represented as:
\begin{equation}\label{Px}
 P_{\overrightarrow{x}}=\prod_{i=1}^d P_{x_i},
 P_{x_i=\{r_i,\varphi_i\}}=\sum_{\alpha=1}^{q} \frac{r_ip_\alpha}{\pi S_\alpha}e^{-\frac{r_i^2+\rho_\alpha^2}{S_\alpha}}\mathrm{I}_0\Big(\frac{2r_i\rho_\alpha}{S_\alpha}\Big)
   \end{equation}
 here the Latin alphabet is used to denote the time-index, and the Greek, the coding level.

In the simplest case of $q=1$ we receive the conventional result:
 \begin{equation}\label{Px0}
 P_{\overrightarrow{x}}=\prod_{i=1}^d \frac{r_i}{\pi S}e^{-\frac{r_i^2}{S}}
   \end{equation}
   which is the Gaussian pdf with zero mean and variance $S$ in polar coordinates. The resulting lower bound is
   \[I_0=\log_2\Big(1+\frac{S}{N(1+C_{nl}S^2)}\Big)\] which is plotted in Fig. \ref{F_C}, where $C_{nl}=6S^2N \sum_{m,n\neq 0}|K_{mn}|^2$, and is in agreement with the asymptotic closed form expression of \cite{Ellis}.

 In case of $q=2$ and for $S>C_{nl}^{-1/2}$, we consider a set of two Gaussian distributions (see Fig. \ref{F_C}b). One is centered around zero power level and it has a fixed variance equal to the maximum peak power of the lower bound with single level $S_1=C_{nl}^{-1/2}$, whereas  the second  one will have the same variance and vanishingly small probability $\delta\rightarrow 0$ and it is centered around a distant power level $\rho^2\gg S_1$. Thus, we consider $p_{\alpha=1,2}=\{1-\delta, \delta\}$, $S_{\alpha=1,2}=\{C_{nl}^{-1/2}, C_{nl}^{-1/2}\}$, and $\rho_{\alpha=1,2}=\{0,\rho\}$. Consequently, the corresponding lower bound on capacity for $\rho^2\gg S_1$ having optimized the free parameter $\delta$ results in the following \emph{analytical bound}
 \[I_1=\log_2\Big(1+\frac{1}{2N\sqrt{C_{nl}}}\Big)+2\sqrt{\pi\rho^2C_{nl}^{1/2}}e^{-\rho^2C_{nl}^{1/2}}\log_2e\]
where the parameter $\rho$ can be found from the power requirement: $2\sqrt{\pi\rho^2C_{nl}^{1/2}}\rho^2e^{-\rho^2C_{nl}^{1/2}}=S-S_1$.
This result shows that, with the considered suboptimal input pdf, we received a monotonically increasing lower capacity bound, which is asymptotically close to the plateau level. Therefore, we proved that signal-noise effects do not decrease capacity with the increase of signal power.  For uncompensated signal-signal distributions without taking into account signal-noise interference a similar result was obtained\cite{Agrell2014}. Further optimization of the pdf can only improve this bound.

To prove this, we consider the ripple distribution for larger number of levels $q$ and with adjustable variances and centers of each distribution. We have found that the numerical optimization results in a monotonically increasing $I_2$  \emph{numerical bound}.

In conclusion, we derived the general finite memory multivariate channel model for describing nonlinear inter-symbol interfering effects in fiber optic communication channels. The model predicts an exponential
decay of the interference with the inter-symbol distance. For the first time we demonstrated that a lower bound on the channel capacity is a monotonically increasing function of signal power. We found an input
signal ripple distribution that enables us to demonstrate this monotonically increasing lower bound on capacity. This provides an information-theoretic tool for estimating the capacity limits of fiber channels, as
well as, for the practical design of efficient compensation algorithms and coding schemes tailored by the nonlinearity.

This work has been supported by the EPSRC
project UNLOC EP/J017582/1.


\begin{thebibliography}{1}
\bibitem{Shannon}  Shannon, C. E.  A Mathematical Theory of Communication, \textit{Bell Syst. Tech. J. } \textbf{27}, 379-423, 623-656 (1948).


\bibitem{Sp}  Splett, A., Kurtzke, C. , \&  Petermann, K. Ultimate transmission capacity of amplified optical fiber communication systems taking into account fiber nonlinearities, in \textit{Tech. Digest of European Conference on Optical Communication} paper MoC2.4. (1993).
\bibitem{Essiambre}  Essiambre, R.-J. \textit{et al.}, Capacity limits of information transport in fiber-optic networks, \textit{Phys. Rev. Lett.} \textbf{101}, 163901 (2008).
\bibitem{GN}   Poggiolini, P. \textit{et al.}, Analytical modeling of non-linear propagation in
uncompensated optical transmission links, \textit{IEEE Photon. Technol. Lett.} \textbf{23}, 742-744 (2011).

\bibitem{IPM}   Djordjevic, I. B.,  Batshon, H. G.,  Xu, L., and  Wang, T., “Coded
polarization-multiplexed iterative polar modulation (PM-IPM) for beyond
400 Gb/s serial optical transmission,” in Proc. Optical Fiber
Communication Conference, Los Angeles, CA, Mar. 2010, p. OMK2


\bibitem{GS1}  Fehenberger, T.,  Alvarado Segovia, A.,  Bocherer, G.,  Hanik, N., “Sensitivity gains by mismatched probabilistic shaping for
optical communication systems,” IEEE Photon. Technol. Lett., vol. 28,
no. 7, pp. 786–789, Apr. 2016.



\bibitem{GS3}     Pan, C.,  Kschischang, F. R., "Probabilistic 16-QAM Shaping in WDM Systems," in Journal of Lightwave Technology (in press).

\bibitem{GS4}  Buchali, F.,  \textit{et al.},
“Rate adaptation and reach increase by probabilistically shaped 64-
QAM: an experimental demonstration,” J. Lightw. Technol., vol. 34,
no. 7, pp. 1599–1609, Apr. 2016.

\bibitem{DJR}  Richardson, D. J. Filling the light pipe, \textit{Science} \textbf{330}, 327–328 (2010).
\bibitem{Radic1}  Temprana, E. \textit{et al.}, Overcoming Kerr-induced capacity limit in optical fiber transmission, \textit{Science}  \textbf{348}, 1445-1448, (2015).
\bibitem{PW}  Winzer, P. J. Scaling Optical Fiber Networks: Challenges \& Solutions,” \textit{Optics \& Photonics News,} \textbf{26} 28–35 (2015).

\bibitem{chapter}  Sorokina, M., Ellis A. \&  Turitsyn S., Optical Information Capacity Processing, chapter in All-Optical Signal Processing, 325-354 (2015).


\bibitem{Agrell2015}  Agrell, E., Durisi, G. \& Johannisson, P. Information-theory-friendly models for fiberoptic
channels: A primer". IEEE Information Theory Workshop  (2015).


\bibitem{Agrell2014}  Agrell,  E.  \textit{et al.},   Capacity of a nonlinear optical channel with finite memory,
\textit{J. Lightwave Technol.} \textbf{16}, 2862-2876 (2014).



\bibitem{D1}  Taghavi, M. H.,  Papen, G. C. \&  Siegel, P. H. On the multiuser capacity
of WDM in anonlinear optical fiber: Coherent communication, \textit{IEEE
Trans. Inf. Theory,}  \textbf{52},  5008–5022 (2006).



\bibitem{D3} Song, H. \&  Brandt-Pearce, M. “A 2-D discrete-time model of physical
impairments in wavelength-division multiplexing systems,” \textit{J. Lightw.
Technol.} \textbf{30}, 713–726 (2012).

\bibitem{Dar}  Dar, R., Shtaif, M. \&  Feder, M. New bounds on the capacity of the
nonlinear fiber-optic channel, \textit{Optics Letters }\textbf{39},  398-401 (2014).


\bibitem{Tao}  Tao, Z. \textit{et al.}, \textit{Journ. Lightwave Technol.} \textbf{33}, 2111-2018 (2015).

\bibitem{Karlsson} Johannisson P. \& Karlsson, M. Perturbation analysis of nonlinear
propagation in a strongly dispersive optical communication system, \textit{J.
Lightw. Technol.,}  \textbf{31},  1273–1282 (2013).

\bibitem{16QAM} W.-R. Peng \textit{et al.}, ECOC 2014, We3.3.4.
\bibitem{quant} Z. Li, W.-R. Peng, F. Zhu, \& Y. Bai, "Optimum quantization of perturbation coefficients for
perturbative fiber nonlinearity mitigation," in \textit{Tech. Digest of European Conference on Optical Communication} paper We.1.3.4. (2014).

\bibitem{ECOC_Essiambre}  Ghazisaeidi, A. \&  Essiambre, R.-J. Calculation of coefficients of perturbative nonlinear pre-compensation for Nyquist pulses, in \textit{Tech. Digest of European Conference on Optical Communication} paper We.1.3.3. (2014).


\bibitem{Ellis}    Rafique D. \& Ellis A. Impact of signal-ASE four-wave mixing on the effectiveness of digital back-propagation in 112 Gb/s PM-QPSK systems, \textit{Opt. Express } \textbf{19}, 3449-3454 (2011).



\bibitem{complexnormal} B. Picinbono,  Second-order complex random vectors and normal distributions, \textit{IEEE Transactions on Signal Processing} \textbf{44}  2637 (1996).


\end{thebibliography}
\end{document}